\documentclass[final, twocolumn, a4paper]{aastex63}
    
\usepackage{color}
\usepackage{enumitem}
\usepackage{hyperref}
\usepackage{verbatim}
\usepackage{amsmath,amstext,mathrsfs}
\usepackage[all]{hypcap} 
\usepackage{afterpage}    
\usepackage{marginnote}
\usepackage{makecell}
\usepackage{subfigure}
\usepackage{multirow}
\usepackage{lineno}

    \shorttitle{magnetic field in Orion A }
    \shortauthors{Zhao et al}
\usepackage{enumitem}
\setlist[enumerate]{listparindent=\parindent}
\makeatletter

\newcommand{\Rmnum}[1]{\expandafter\slowromancap\romannumeral #1@}
\makeatother

\begin{document}
    \hfuzz = 150pt
\title{\Large\bfseries  Magnetic Field of Molecular Gas Measured with the Velocity Gradient Technique  I. Orion\,A}
\correspondingauthor{Mengke Zhao,Jianjun Zhou}
\email{zhaomengke@xao.ac.cn,zhoujj@xao.ac.cn}

\author[0000-0003-0596-6608]{Mengke Zhao}
\affil{Xinjiang Astronomical Observatory, Chinese Academy of Sciences,  Urumqi, 830011, People's Republic of China}
\affil{University of Chinese Academy of Sciences, Beijing, 100049, People's Republic of China}
\author[0000-0003-0356-818X]{Jianjun Zhou}
\affil{Xinjiang Astronomical Observatory, Chinese Academy of Sciences,  Urumqi, 830011, People's Republic of China}
\affil{Key Laboratory of Radio Astronomy, Chinese Academy of Sciences  Urumqi,830011, People's Republic of China}

\author[0000-0002-8455-0805]{Yue Hu}
\affiliation{Department of Physics, University of Wisconsin-Madison, Madison, WI 53706, USA}
\affiliation{Department of Astronomy, University of Wisconsin-Madison, Madison, WI 53706, USA}
\author{A. Lazarian}
\affiliation{Department of Astronomy, University of Wisconsin-Madison, Madison, WI 53706, USA}
\affiliation{Centro de Investigación en Astronomía, Universidad Bernardo O’Higgins, Santiago, General Gana 1760, 8370993, Chile}

\author[0000-0002-4154-4309]{Xindi Tang}
\affil{Xinjiang Astronomical Observatory, Chinese Academy of Sciences,  Urumqi, 830011, People's Republic of China}
\affil{Key Laboratory of Radio Astronomy, Chinese Academy of Sciences  Urumqi,830011, People's Republic of China}
\author[0000-0003-3389-6838]{Willem A. Baan}
\affil{Xinjiang Astronomical Observatory, Chinese Academy of Sciences,  Urumqi, 830011, People's Republic of China}
\affil{Netherlands Institute for Radio Astronomy ASTRON, 79901 PD Dwingeloo, the Netherlands}
\author{Jarken Esimbek}
\affil{Xinjiang Astronomical Observatory, Chinese Academy of Sciences,  Urumqi, 830011, People's Republic of China}
\affil{Key Laboratory of Radio Astronomy, Chinese Academy of Sciences  Urumqi,830011, People's Republic of China}
\author[0000-0002-8760-8988]{Yuxin He}
\affil{Xinjiang Astronomical Observatory, Chinese Academy of Sciences,  Urumqi, 830011, People's Republic of China}
\affil{Key Laboratory of Radio Astronomy, Chinese Academy of Sciences  Urumqi,830011, People's Republic of China}
\author{Dalei Li}
\affil{Xinjiang Astronomical Observatory, Chinese Academy of Sciences,  Urumqi, 830011, People's Republic of China}
\affil{Key Laboratory of Radio Astronomy, Chinese Academy of Sciences  Urumqi,830011, People's Republic of China}
\author{Weiguang Ji}
\affil{Xinjiang Astronomical Observatory, Chinese Academy of Sciences,  Urumqi, 830011, People's Republic of China}]
\author{Kadirya Tursun}
\affil{Xinjiang Astronomical Observatory, Chinese Academy of Sciences,  Urumqi, 830011, People's Republic of China}
\affil{Key Laboratory of Radio Astronomy, Chinese Academy of Sciences  Urumqi,830011, People's Republic of China}

\begin{abstract}

Magnetic fields play an important role in the evolution of molecular clouds and star formation. 
Using the Velocity Gradient Technique\,(VGT) model, we measured the magnetic field in Orion\,A using the $^{12}$CO, $^{13}$CO, and C$^{18}$O\,(1-0) emission lines at a scale of $\sim$ 0.07\,pc. 
The measured B-field shows an east-west orientation that is perpendicular to the integral shaped filament of Orion\,A at large scale. 
The VGT magnetic fields obtained from $^{13}$CO and C$^{18}$O are in agreement with the B-field that is measured from the Planck 353\,GHz dust polarization at a scale of $\sim$ 0.55\,pc. 
Removal of density effects by using a Velocity Decomposition Algorithm can significantly improve the accuracy of the VGT in tracing magnetic fields with the $^{12}$CO\,(1-0) line. 
The magnetic field strength of seven sub-clouds, OMC-1, OMC-2, OMC-3, OMC-4, OMC-5, L\,1641-N and NGC\,1999 has also been estimated with the Davis-Chandrasekhar–Fermi (DCF) and MM2 technique, and these are found to be in agreement with previous results obtained from dust polarization at far-infrared and sub-millimeter wavelengths. 
At smaller scales, the VGT prove a good method to measure magnetic fields.

\end{abstract}

\keywords{Interstellar medium (847); Interstellar magnetic fields (845); Interstellar dynamics (839)}

\section{Introduction}

Magnetic fields play a key role in regulating the formation of molecular clouds and their evolution \citep{1981MNRAS.194..809L,2015MNRAS.452.2410S,2003ApJ...585..850M,2007ARA&A..45..565M}. 
However, their role in the star-formation process is not entirely understood \citep{2015Natur.520..518L,2012ARA&A..50...29C}. 
In addition, turbulence effects are considered to be another key factor affecting the dynamics of star formation processes in molecular clouds. 
Together with gas self-gravity, these processes appear at all physical scales and at different evolutionary stages \citep{1965ApJ...142..584P,1979cmft.book.....P,1966ApJ...146..480J,2011Natur.479..499L,2013ApJ...768..159H,2014ApJ...783...91C,2015ARA&A..53..501A}. 
In addition to the polarization measurements, the Velocity Gradient Technique (VGT; \citealt{2017ApJ...835...41G,2017ApJ...837L..24Y,2018ApJ...853...96L,2018MNRAS.480.1333H}) provides a new approach to study interstellar magnetic fields. 
The velocity gradients of these elongated eddies are expected to be perpendicular to the local B field orientation.
Different from the synchrotron or dust polarization studies, the VGT explores the anisotropy of the magneto-hydrodynamic (MHD) turbulence, which means that turbulent eddies are elongated along the percolated magnetic field lines \citep{1995ApJ...438..763G,1999ApJ...517..700L}. 
This method has been applied successfully to several molecular clouds \citep{2019NatAs...3..776H,2021ApJ...912....2H}.

Strong gravitational collapse will break the perpendicular relative orientation of the velocity gradient and the magnetic fields. 
Because the gravitational force pulls the plasma in the direction parallel to the magnetic field and produces a most significant acceleration, the velocity gradients in this location are dominated by the gravitational acceleration being parallel to the magnetic fields. 
This phenomenon of domination by gravitational acceleration has indeed been detected in the Serpens molecular clouds \citep{2019NatAs...3..776H,2021ApJ...912....2H}, G34.43+00.24 \citep{2019ApJ...878...10T}, and NGC 1333 \citep{2021MNRAS.502.1768H}. 
This reaction to self-gravity enables the VGT to reveal self-gravity-dominated regions, as well as quiescent areas supported by turbulence, thermal pressure, and magnetic fields.
The velocity information of the molecular gas is obtained from Doppler-shifted spectral lines. Because of the effect of a velocity caustic \citep{2000ApJ...537..720L}, the observed intensity distribution in a given velocity channel is defined by both the emitter density and the velocity distributions. 
To separate the density and velocity contributions, \cite{2021ApJ...910..161Y} proposed a new method, i.e., the Velocity Decomposition Algorithm (VDA). 
The accuracy of the VGT is expected to be improved by eliminating the dependence on density. 

In this work, we target the Orion A molecular cloud, which is the nearest high-mass star-forming region located at a distance of around 400 pc \citep{2007A&A...474..515M,2017ApJ...834..142K}. 
The complex structure of the Orion\,A filament can be seen from H$_2$ column density ditribution in Orion\,A (see Figure\,\ref{fig1}).
Abundant polarization observations at different wavelengths have been performed for this source, including near-infrared (NIR) polarimetry \citep{2011ApJ...741..112P}, far-infrared polarimetry \citep{1998ApJ...493..811S,2019ApJ...872..187C,2018JAI.....740008H}, and sub-millimeter polarimetry \citep{1998ApJ...493..811S,2010ApJ...717.1262T,2017ApJ...842...66W,2017ApJ...846..122P,2021MNRAS.503.3414P,2020A&A...641A..12P,2020A&A...641A...1P}. 
Recently, the CARMA-NRO Orion Survey \citep{2018ApJS..236...25K} provided high-resolution $^{12}$CO, $^{13}$CO, and C$^{18}$O\,(1-0) spectral line data (beam size $\sim$ 6$''$-10$''$), which is from CARMA observations combined with single-dish data from the Nobeyama telescope. 
In the following, these high-resolution CO data may make it possible to generate both a multi-scale (from 10 to 0.1 pc) and multi-wavelength view of the magnetic field as it interacts with the multi-phase gas in Orion A. 

In this work, we aim to measure the magnetic field structure of Orion\,A with VGT method using $^{12}$CO, $^{13}$CO, and C$^{18}$O\,(1-0) spectral line data.
The paper is organized as follows. 
In Sect.\,\ref{sec2}, we provide the details of the observational data used in this work. 
In Sect.\,\ref{sec3}, we describe the details of the VGT method, and the VDA algorithms.
In Sect.\,\ref{result}, the magnetic field measured with VGT and VGT-VDA method have been described.
Sect.\,\ref{discussion} will show the realm of applicability of magnetic field measurements with VGT and some physical parameters. 
A summary has been provided in Sect.\,\ref{sum}.

\section{Archival Data}\label{sec2}

\subsection{The CO emission data}

Carbon monoxide is a ideal tracer of the kinematic characteristics of molecular clouds. 
There are three resolution spectral lines $^{12}$CO\,(1-0), $^{13}$CO\,(1-0) and C$^{18}$O\,(1-0) from  CARMA-NRO Orion Survey \citep{2018ApJS..236...25K}, where CARMA observations were combined with single-dish data from the Nobeyama 45 m telescope to provide extended images at about 0.01 pc resolution. 
The final maps have an angular resolution of about 8$''$ (from 6$''$ to 10$''$) and a pixel size of 2$''$. 
The velocity resolution is 0.25 km s$^{-1}$ for $^{12}$CO and 0.22 km s$^{-1}$ for $^{13}$CO and C$^{18}$O.

\begin{figure}
    \centering
    \includegraphics[width=8.5cm]{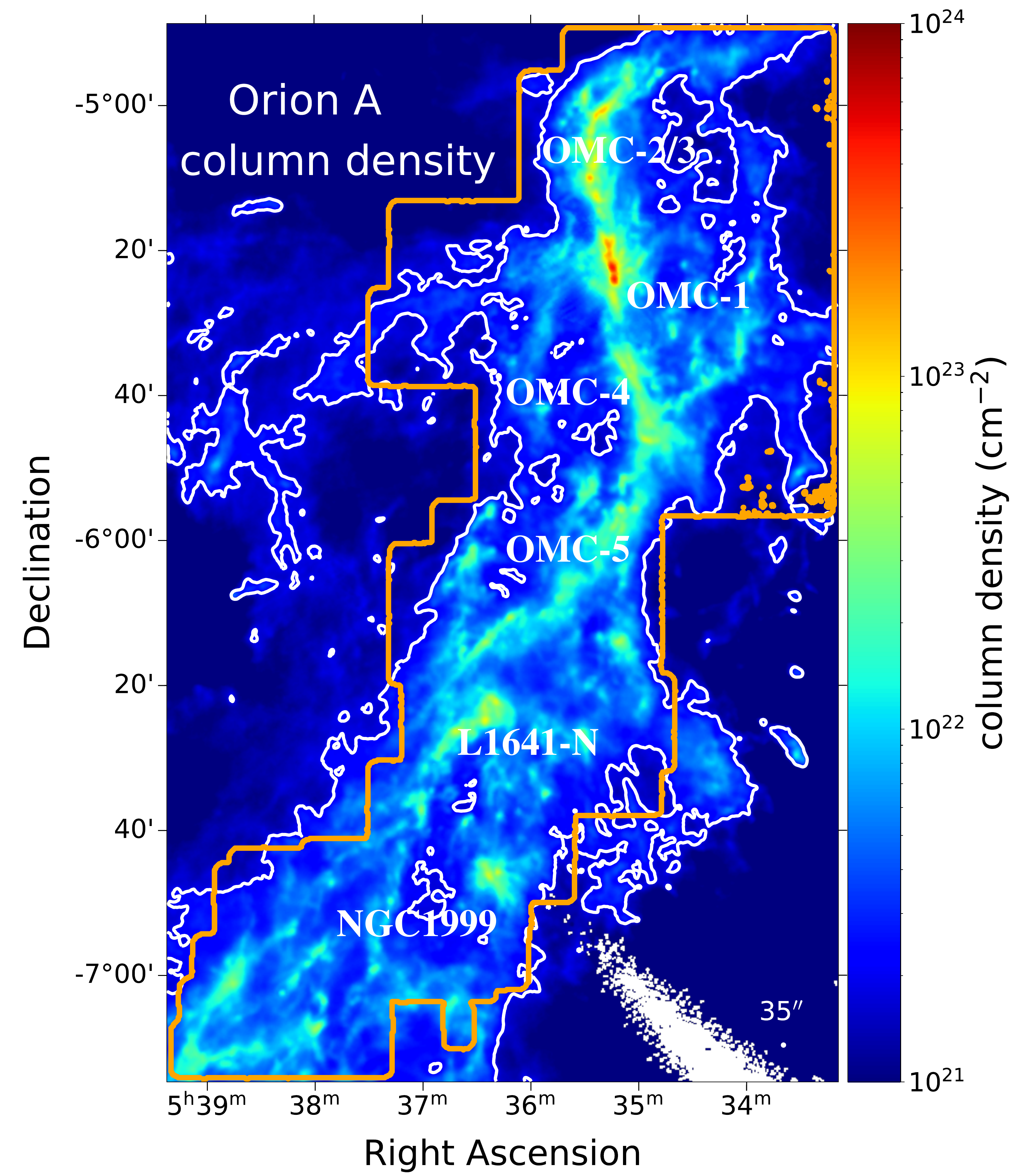}
    \caption{Distribution of H$_2$ column density derived from Herschel continuum data in Orion A \citep{2010A&A...518L...2P,2010A&A...518L...3G,2013ApJ...763...55R,2013ApJ...777L..33P}.
    The white contour level represents 3 $\times$ $10^{21}$ cm$^{-2}$. The orange contours show the region covered by the CARMA-NRO Orion Survey \citep{2018ApJS..236...25K}. }
    \label{fig1}
\end{figure}

\subsection{The Polarization data}
The Planck satellite\footnote{\url{http://www.esa.int/Planck}} provides the 353\,GHz thermal dust polarized emission \citep{2020A&A...641A..12P,2020A&A...641A..11P} tracing the large scale magnetic field in Orion\,A. 
Based on the observations of the High-Frequency Instrument (HFI) at 353 GHz, the Stokes parameters I, Q, U and their dispersion value ($\sigma \rm  I$, $\sigma \rm Q$, $\sigma \rm U$) maps have been obtained. 
The resolution of these maps is 5$'$ and the pixel size is $\sim$ 1.71$'$. 
The polarization angle from the HFI Stokes maps may be calculated as:
\begin{equation}
    \begin{aligned}
        \rm \psi_{Planck} = 0.5 \times arctan(U,Q) \,,
    \end{aligned}
\end{equation}
where $\psi_{\rm Planck}$ varies from -90$^\circ$ to 90$^\circ$ with the HEALPix convention. One has to use $\psi$=0.5 $\times$ arctan(-U, Q) to convert the Planck measurement to this IAU convention. 
The magnetic field (hereafter, the B-field) orientation can be obtained by adding 90$^\circ$ to the polarization angle: $\psi_{\rm B}$\,=\,$\psi_{\rm Planck}$\,+\,90$^\circ$.

Furthermore, the B-field orientation in equatorial coordinates (FK5, J2000) is obtained using the following angle relation \citep{1998MNRAS.297..617C}:
 \begin{equation}
     \begin{aligned}
        \rm \psi_r= arctan\big[\frac{cos(\emph{l} - 32.9^\circ)}{cos\emph{b}\,cot62.9^\circ - sin\emph{b}\,sin(\emph{l} - 32.9^\circ)}\big] \,,
     \end{aligned}
 \end{equation}
where $\psi_{\rm r}$ is the angle relation for spherical triangles between Equatorial and Galactic coordinate systems. 
$l$ and $b$ show the pixel position information of the Galactic coordinates. 
The magnetic field orientation is then transformed from  Galactic ($\theta_{\rm GPA}$) to  Equatorial ($\theta_{\rm EPA}$) coordinates by:
 \begin{equation}
     \begin{aligned}
        \rm\theta_{EPA} = \rm\theta_{GPA} - \rm\psi_r \,,
     \end{aligned}
 \end{equation}
 

\section{method}\label{sec3}
\subsection{The Velocity Gradient Technique}\label{VGTm}
The Velocity Gradient Technique (VGT; \citealt{2017ApJ...835...41G,2018ApJ...853...96L,2018MNRAS.480.1333H}) is the main analysis tool used in this work and has been developed on the basis of the anisotropy of magneto-hydrodynamic turbulence \citep{1995ApJ...438..763G} and fast turbulent reconnection theories \citep{1999ApJ...517..700L}. 
We use the Velocity Channel Gradients (VChGs) as the main model of VGT (after here short as VGT).
For extracting the velocity information from Position-Position-Velocity PPV cubes, thin velocity channels Ch(x,y) were employed. The gradient map $\psi^{s}_{g}$ is then calculated by:


\begin{equation}
    \begin{aligned}
        \rm\bigtriangledown_x Ch_i(x,y) = Ch_i(x,y) - Ch_i(x-1,y) \,,
    \end{aligned}
\end{equation}
\begin{equation}
    \begin{aligned}
        \rm\bigtriangledown_y Ch_i(x,y) = Ch_i(x,y) -Ch_i(x,y-1)\,, 
    \end{aligned}
\end{equation}
\begin{equation}\label{eq6}
    \begin{aligned}
        \rm\psi^i_g = tan^{-1} (\frac{\bigtriangledown_y Ch_i(x,y)}{\bigtriangledown_x Ch_i(x,y)} \,,
    \end{aligned}
\end{equation}
where $\bigtriangledown_x$Ch$_i$(x,y) and $\bigtriangledown_y$Ch$_i$(x,y) are the x and y components of the gradient, respectively.
This is done for all pixels with spectral line emission having a signal-to-noise ratio greater than 3.

The orientation of the magnetic field is found to be perpendicular to the velocity gradient, as long as these gradients are statistically significant. 
A sub-block averaging method \citep{2017ApJ...837L..24Y}  has been used to export the velocity gradients from the raw gradients within a sub-block of interest and then to plot the corresponding histogram. 
Gradients for each channel are then calculated by adaptive sub-block averaging, which results in eigen-gradient maps $\psi^i_{gs}(x,y)$ with i = 1,2,...,$n_v$. 
Pseudo-Stokes-parameters $Q_g$ and $U_g$ of the gradient-inferred magnetic field may then be constructed by:
\begin{equation}
    \begin{aligned}
        \rm Q_g(x,y) = \sum\limits_{i=1}^{n_v} I_i(x,y)cos(2\psi^i_{gs}(x,y))  \,,
    \end{aligned}
\end{equation}
\begin{equation}
    \begin{aligned}
        \rm U_g(x,y) = \sum\limits_{i=1}^{n_v} I_i(x,y)sin(2\psi^i_{gs}(x,y)) \,,
    \end{aligned}
\end{equation}
\begin{equation}
    \begin{aligned}
        \rm\psi_{g} = \frac{1}{2}tan^{-1}\frac{U_{g}}{Q_{g}} \,,
    \end{aligned}
\end{equation}
where $\psi_g$ is the pseudo polarization angle. 
This pseudo polarization angle is perpendicular to the POS orientation angle of the magnetic field: $\psi_{\rm B} = \psi_{\rm g} + \pi /2$ .

For the dense region, the turbulent flow will be modified thoroughly by self-gravity and the Velocity gradient orientation will change from perpendicular to parallel to magnetic field \citep{2017arXiv170303026Y,2020ApJ...897..123H}. 
Therefore the velocity gradient orientation angle in gravity-dominated region will be re-rotate 90 degrees.
It is calculated by:
\begin{equation}\label{eq10}
    \begin{aligned}
        \rm\psi_B^S = \psi_B \, +\,\pi/2,
    \end{aligned}
\end{equation}
where $\psi_B^S$ is VGT orientation angle in the case of self-gravity and $\psi_B$ is the pseudo magnetic field angle measured by VGT in the case of turbulence dominated regions.
The orientation of the turbulence velocity gradient will be parallel to the magnetic field orientation.

\subsection{Velocity Decomposition Algorithm}
The Velocity decomposition algorithm (VDA, \citealt{2021ApJ...910..161Y}) is a new method to separate velocity and density fluctuations from a PPV cube using its statistical properties \citep{2000ApJ...537..720L}. 
The sonic Mach number in star formation regions is usually greater than unity, and the supersonic version of the VDA algorithm makes it possible to only obtain the velocity flux structure from a PPV cube using the thin channel formulation of the VGT method.
The VDA allows a separation of the pure velocity caustics from the PPV cube. 
In theory, the VGT orientation applied the VDA method (after here, VGT-VDA) would be closer to the plasma motion direction and better trace the local magnetic field. 

The supersonic VDA algorithm to get the pure velocity caustics structures \citep{2000ApJ...537..720L} for the each channel of PPV cube is based on the following expression:
\begin{equation}
    \begin{aligned}
        \rm V(X, v, \Delta v) = - c^2_s \frac{\partial Ch(X, v, \Delta v)}{\partial v} \,,
    \end{aligned}
\end{equation}
where Ch(X, v, $\Delta$v) is the channel of the PPV cube, X means the position, v is the local velocity, and $\Delta$v is the velocity channel width. $c_s$ can be calculated by assuming a uniform temperature $\sim$ 10 K, which results in a value $c_s$ $\sim$ 186 m s$^{-1}$.
When using the pseudo PPV cube V(X, v, $\Delta$v) with only the velocity contribution, the raw gradients $\psi^i_g$ (see Eq.\,\ref{eq6}) may be re-applied with VGT to improve the accuracy to trace the magnetic field.

The use of this technique requires high-SNR spectral line data \citep{2021ApJ...910..161Y}. 
This technique has not yet been applied on self-gravity regions, and we will use the CO data in Orion A to see if applying VDA improves the accuracy of VGT to trace magnetic field.

\begin{figure*}
    \centering
    \includegraphics[width = 18cm]{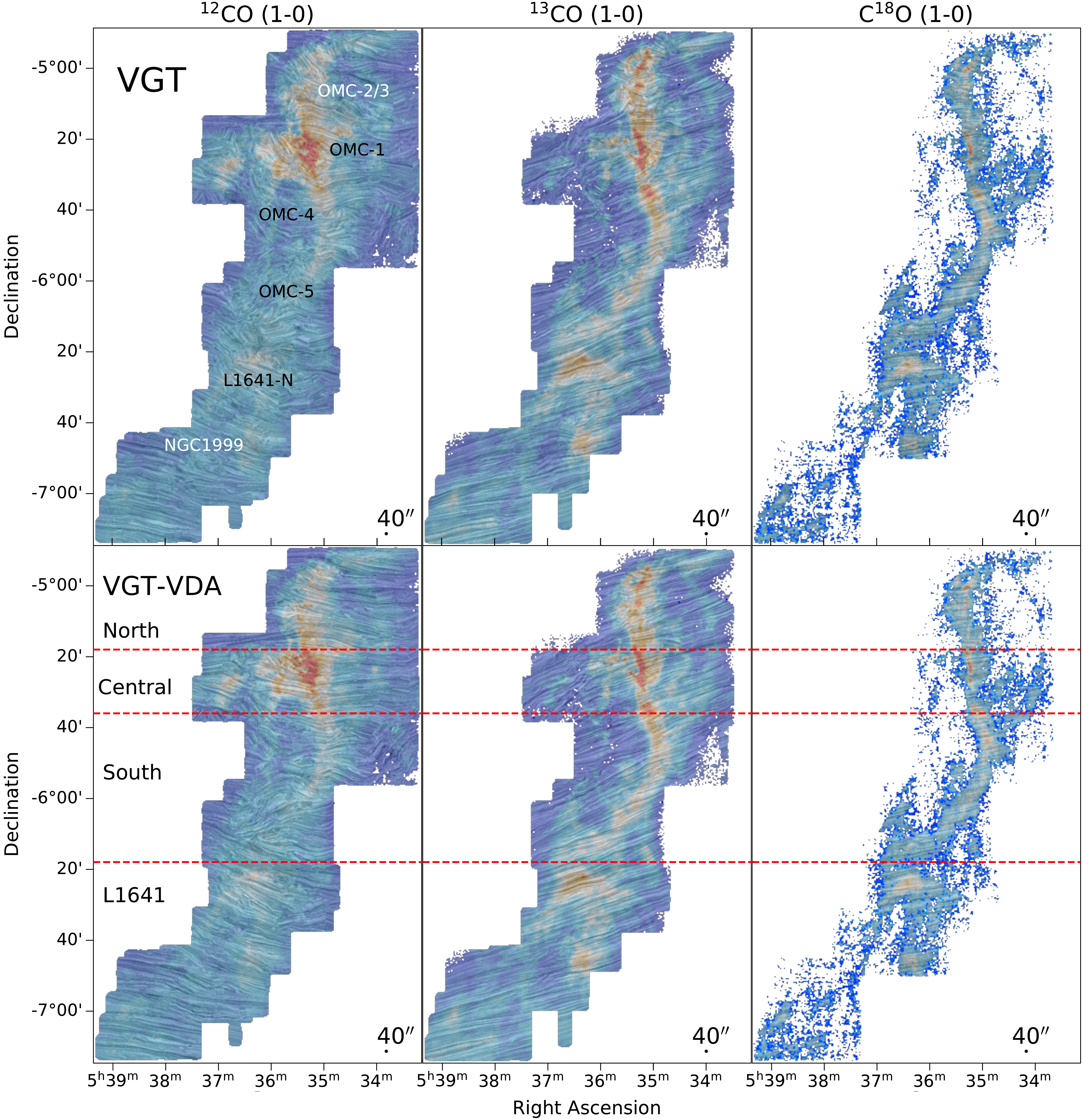}
    \caption{Three line integral convolution \citep[LIC]{Cabral_lic} maps from VGT using the $^{12}$CO\,(left), $^{13}$CO\,(middle), C$^{18}$O\,(right) emission. 
    The beam size has been shown in the lower right corner of each panel. 
    The top panels present the magnetic field morphology measured with the VGT method. 
    The bottom panels are the B-field distributions measured with the combined VGT-VDA method. 
    The colour backgrounds are the intensity maps for $^{12}$CO, $^{13}$CO and C$^{18}$O\,(1-0) integrated from $V_{\rm LSR}$ = 1 to 15 km\,s$^{-1}$.}
    \label{fig2}
\end{figure*}



\section{Results}\label{result}

\subsection{Magnetic Fields measured with VGT}\label{s4.1}

The VGT treats the regions dominated by turbulence or self-gravity differently and it is important to find out which region in Orion\,A favor the one or the other scenario. 
The column density probability function (N-DPFs) \citep{2011MNRAS.416.1436B,2018ApJ...863..118B,2019MNRAS.482.5233K} provides a simple way to make this distinction. 
The N-DPFs follow a log-normal\,(LN) distribution in the case of turbulence dominated regions, but they will follow a power-law\,(PL) distribution in self-gravity-dominated regions \citep{2008ApJ...680.1083R,2018ApJ...863..118B,2019MNRAS.482.5233K}. 
The transition point of this LN-PL model is a key for distinguishing the VGT model that is dominated by turbulence or self-gravity. 
Recent studies shows that the critical column density at this transition point is $\sim 3\times$ $10^{21}$ cm$^{-2}$ \citep{2021A&A...653A..63S}, i.e. if the column density of one region is greater than this density, self-gravity could be dominating. 
This dense region in Orion\,A includes the large integral shaped filament (hereafter, ISF), dense clumps, L1641-N and NGC1999, and the gas around them which has high column density N(H$_2$) ($\textgreater$ 3\,$\times$\,10$^{21}$ cm$^{-2}$).
Assume that Orion\,A is a long cylinder, the volume density of N-DPFs' transition point (column density $\sim$ 3\,$\times$\,10$^{21}$ cm$^{-2}$; \citealt{2021A&A...653A..63S}) is around $\times$\,10$^3$ cm$^{-3}$ by estimating the effective radius of cloud ($\sim$ 0.61\,pc, the details see Appendix.\ref{A-A}).
This result is in agreement with that from \citealt{2021ApJ...912....2H}, in which the VGT method reveals that the self-gravity is occurring at volume density n$_0$ $\geq$ 10$^3$\,cm$^{-3}$.
However, self-gravity could be more localized to the clumps and cores while the envelope may be more diffuse. 
There is a possibility of a global gravitational contraction \citep{1981MNRAS.194..809L,2011MNRAS.416.1436B,2013ApJ...779..185K,2019MNRAS.490.3061V} that could cause an inflow \citep{2020ApJ...897..123H} and change the direction of the velocity gradient.
This region of this work in Orion A is dominated by self-gravity where the gravitationally collapse could occur at the core’s scale and gravitationally contraction could occur at the cloud’s scale.
The distribution of the H$_2$ column density in Fig.\,\ref{fig1} suggests that the whole region covered by the CO lines \citep{2018ApJS..236...25K} in Orion A has high column density ($\textgreater$ 3\,$\times$\,10$^{21}$ cm$^{-2}$) and could dominated by self-gravity. 
One thing to note is that the power-law DPF model remains a statistical concept. 
When plotting N-DPFs of the overall region, it is possible that some small sub-regions are dominated by turbulence and are overwhelmed.


Several methods have been applied to measure the magnetic field morphology in Orion\,A. 
The magnetic field measured with VGT is called pseudo magnetic field after here.
Six pseudo magnetic field line integral convolution\,\citep[LIC]{Cabral_lic} maps are shown in Fig.\,\ref{fig2}. 
The top panels in this Fig.\,\ref{fig2} show the pseudo magnetic field orientation LIC maps measured by the VGT method using the $^{12}$CO, $^{13}$CO and C$^{18}$O\,(1-0) spectral lines.
The velocity range of the three CO emissions has been set to [1,\,15] km s$^{-1}$. 
Sub-block averaging was used to determine the pseudo beam of the VGT results (see Sect.\,\ref{VGTm}) using a Sub-block size set as 20$\times$20\, pixels. 
The resolution of the pseudo B-field would be accessing 40$''$ ($\sim$ 0.07\,pc). where the region is dominated by self-gravity, the VGT angle would be re-rotated by 90$^\circ$ again (see Eq\,\ref{eq10}).

The VGT explores the anisotropy (the elongation) of the turbulent eddies resulting from magneto-hydrodynamic (MHD) turbulence in the presence of magnetic fields under sub-Alfvénic and supersonic conditions \citep{2006ApJ...645L..25L}.
However, under supersonic conditions, the presence of shocks and also of self-gravity, the MHD turbulent anisotropies will be affected by the local density \citep{2021ApJ...910..161Y}.
This effect is most serious for higher density regions, where the density contribution to the thin velocity channels could degrade the accuracy of VGT.  
Under these conditions, the "velocity decomposition algorithm" (VDA) may be used to separate the velocity and density contributions from the PPV (position-position-velocity) cube \citep{2021ApJ...910..161Y}.  
The density contribution may then be removed from the PPV cube and the VGT-VDA methods may more accurately re-measure the turbulence related velocity fluctuations. 
The physical properties from the MHD turbulent theory \citep{1995ApJ...438..763G} and the statistics from PPV cubes \citep{2000ApJ...537..720L} have been predicted well by velocity caustics\citep{2000ApJ...537..720L}. 
These velocity caustics are important for the VGT method, in particular for the supersonic case where there is density contamination. 
Therefore, the removal of density effects by the VDA method will enhance velocity caustics and improve the accuracy of the VGT method. 

Considering that Orion\,A is a dense region, its magnetic fields should be studied using the combined VGT-VDA method. 
The result of this study using the same three molecular tracers ($^{12}$CO,$^{13}$CO and C$^{18}$O (J = 1-0)) has been shown in Fig.\,\ref{fig2} in the bottom panels. 
The general orientation of the pseudo magnetic field measured by VGT-VDA is similar for tracers and is perpendicular to the ISF and reveals more details in the B-field LIC maps.

\begin{figure*}
    \centering
    \includegraphics[width = 18.5cm]{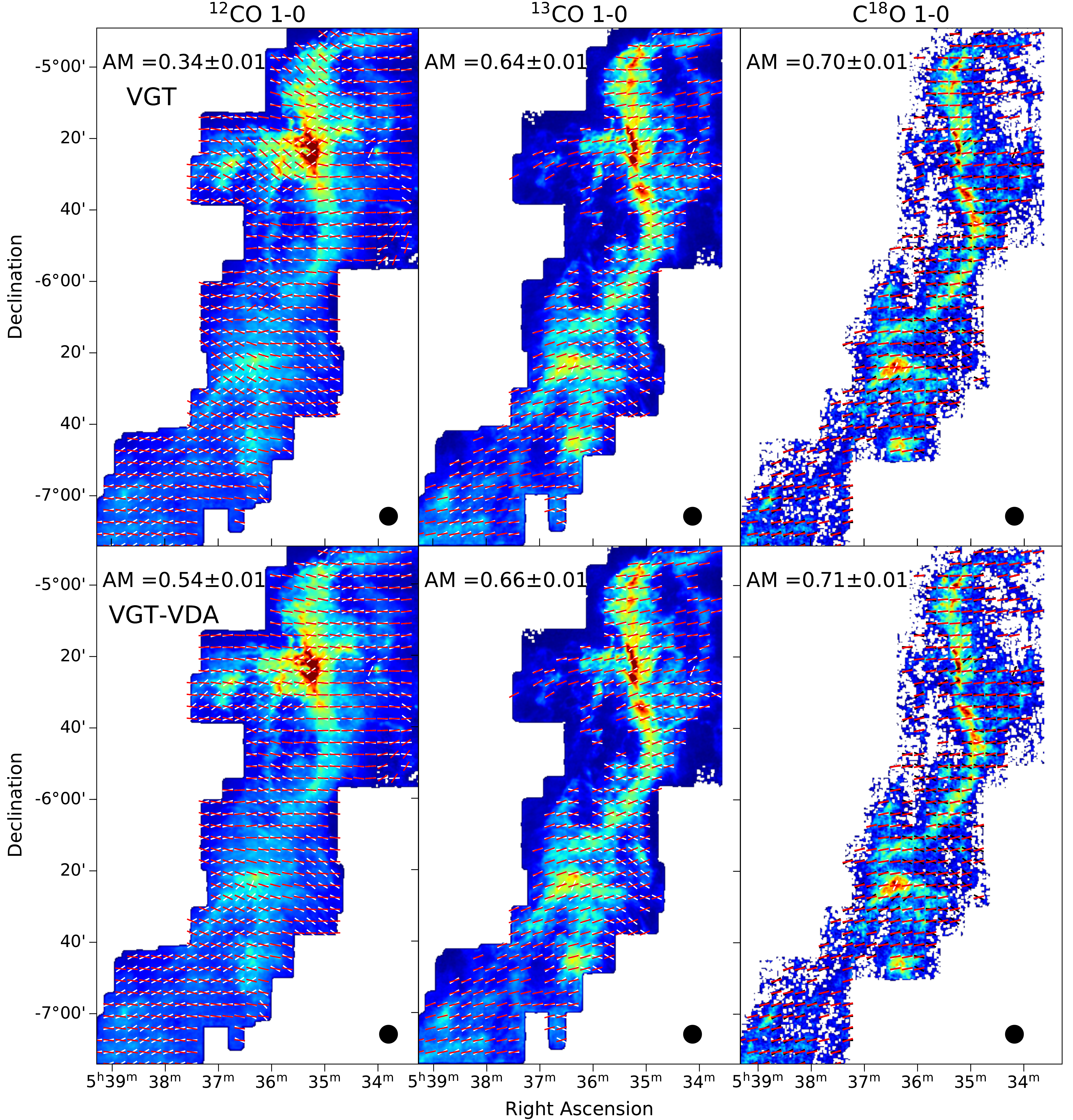}
    \caption{The magnetic fields using the CO data and the Planck HFI data. 
    Top: results of magnetic field orientation from Planck polarization (white vectors) and CO VGT (red vectors) in Orion A. 
    Bottom: White and black vectors are for the Planck polarization and the red vectors show the magnetic field directions derived from VGT-VDA method for three molecular tracers  $^{12}$CO 1-0 (left), $^{13}$CO 1-0 (middle) and C$^{18}$O 1-0 (right). 
    The beam size has been shown in the bottom right corners of each panel. 
    The background are same as Fig.\,\ref{fig2}. 
    The mean AM values and their uncertainties are shown in the top left corner of each panel.}
    \label{VGT/P}
\end{figure*}

\subsection{Magnetic Field Morphology}

The large Integral Shaped Filament (ISF, \citealt{1987ApJ...312L..45B}) is one of the most impressive structural features in Orion\,A, which includes regions OMC-1,\,2,\,3,\,4,\,5 (see Fig.\,\ref{fig2}). 
The pseudo magnetic field orientations from two measuring methods (VGT, and VGT-VDA) are generally similar. 
The pseudo magnetic field directions are almost perpendicular to the long axis of ISF and may vary in individual sub-regions. 
For instance, the pseudo magnetic field in the vicinity of OMC-1 is more scattered, while in the dense region L\,1641N the magnetic field directions point towards the center of its density clumps. 
The Orion\,A generally can be separated into four sub-regions as shown in Fig.\,\ref{fig2}, i.e., the  Orion Nebula Cluster components ONC-North, ONC-Central ONC-South, and L\,1641N. 
In the following, the pseudo magnetic field morphology measured in those sub-regions with VGT and VGT-VDA is displayed in Fig.\,\ref{VGT/P} and is described below in detail.

\noindent\textbf{\textcolor{blue}{ONC-North}} - This region includes sub-regions OMC-2 and OMC-3. 
The pseudo magnetic field orientation in this north-south ISF region is perpendicular to this filament. 
The pseudo B-field from $^{13}$CO VGT is similar to that from C$^{18}$O VGT.
Comparing with pseudo B-field orientations derived from $^{13}$CO and C$^{18}$O, the result obtained from $^{12}$CO has an offset.
$^{12}$CO traces diffuse gas and $^{13}$CO and C$^{18}$O probe dense gas, which could lead slightly different pseudo B-field derived from $^{12}$CO, $^{13}$CO, and C$^{18}$O lines.
Its pseudo B-field orientation tends to be close to a northeast-southwest direction. 

\noindent\textbf{\textcolor{blue}{ONC-Central}} - This is the main sub-region OMC-1. 
The pseudo magnetic field orientation by VGT for $^{13}$CO and C$^{18}$O is perpendicular to the ISF filament. 
However, at the Orion-Bar the orientation of the B-field for $^{12}$CO, $^{13}$CO, and C$^{18}$O is parallel to the structure. 
At east of OMC-1, the pseudo magnetic field shows a disturbance at a cavity structure in the nearby Pillars region \citep{2018ApJS..236...25K}.

\noindent\textbf{\textcolor{blue}{ONC-South}} - 
At ISF filament sub-regions OMC-4 and OMC-5, the pseudo magnetic field orientation is perpendicular to the filament direction.
The pseudo B-field orientations from $^{13}$CO and C$^{18}$O are nearly the same, while the morphology from $^{12}$CO shows a relatively complex structure. 
At other diffuse regions, the general orientation of the B-field is east-west direction and again there are some disturbances around the Pillars region \citep{2018ApJS..236...25K}.

\noindent\textbf{\textcolor{blue}{L1641}} - At the dense region L\,1641N, the pseudo B-field orientations distribute along the east-west direction. 
This region shows two sub-filaments in the form of an inverted 'V' and the pseudo magnetic field orientations from the three CO tracers are perpendicular to these integrated intensity distributions.


\section{Discussion}\label{discussion}

\subsection{A Comparison of the B-field Derived with VGT from different CO molecules.}

Figure\,\ref{fig2} shows that the pseudo magnetic field morphology measured with VGT from $^{13}$CO is similar to that from C$^{18}$O. 
The pseudo magnetic field directions are mainly east-west and are perpendicular to the shape of the ISF. 
In some diffuse regions, the distorted magnetic field follows the density structure. 
East of ONC-Central, the pseudo B-field orientations derived with VGT from $^{13}$CO follow the shape of the gas range. 
This is not evident for VGT with C$^{18}$O.

The pseudo magnetic field morphology measured from VGT for $^{12}$CO is similar to that for $^{13}$CO and C$^{18}$O in dense regions and different in diffuse regions. 
At the ISF, the pseudo B-field orientations from the VGT for $^{12}$CO are perpendicular to the ISF shape and similar to that from the VGT for $^{13}$CO and C$^{18}$O. 
At the east side of ONC-Central, the pseudo B-field orientations derived from the VGT for $^{12}$CO show a more obvious distribution along the dense structure than that from VGT for $^{13}$CO. 
At the L1641 region, the pseudo B-field directions from the VGT for $^{12}$CO are close to the northwest-southeast direction rather than the southwest-northeast direction found from the VGT for $^{13}$CO.

The $^{12}$CO, $^{13}$CO, and C$^{18}$O have different optical depths as the regions traced by $^{12}$CO are more diffuse and closer to the surface of the molecular structures than those traced by $^{13}$CO, and C$^{18}$O. 
Therefore, also the VGT for $^{12}$CO is a good tracer for the magnetic field in more diffuse regions and the VGT for $^{13}$CO and C$^{18}$O would trace the magnetic field in dense regions.

\subsection{Comparison of B-field Derived from VGT and Dust Polarization}

In order to compare the magnetic field orientation obtained with different measuring methods, the offset angle $\theta_r$ between the dust B-field polarization $\phi_B$ and the pseudo magnetic field orientation of VGT $\psi_B^S$ is defined as:
\begin{equation}\label{off}
    \begin{aligned}
        \rm\theta_r = |\phi_B - \psi_B^S| \,,
    \end{aligned}
\end{equation}
The offset angle $\theta_r$ is then quantified by the Alignment Measure (AM, \citealt{2017ApJ...835...41G}) defined as:
\begin{equation}\label{AMeq}
    \begin{aligned}
        \rm AM = 2(\left \langle cos^2\theta_r \right \rangle - \frac{1}{2}) \,.
    \end{aligned}
\end{equation}
The range of AM values would be from -1 to 1, where  AM values close to 1 means that $\phi_B$ is parallel to$\psi_B^S$ and an AM value close to -1 indicates that $\phi_B$ is perpendicular to $\psi_B^S$. 
The uncertainty in the AM value, $\sigma_{AM}$, may be given by a standard deviation divided by the square root of the sample size.

\begin{figure*}
    \centering
    \includegraphics[width=18cm]{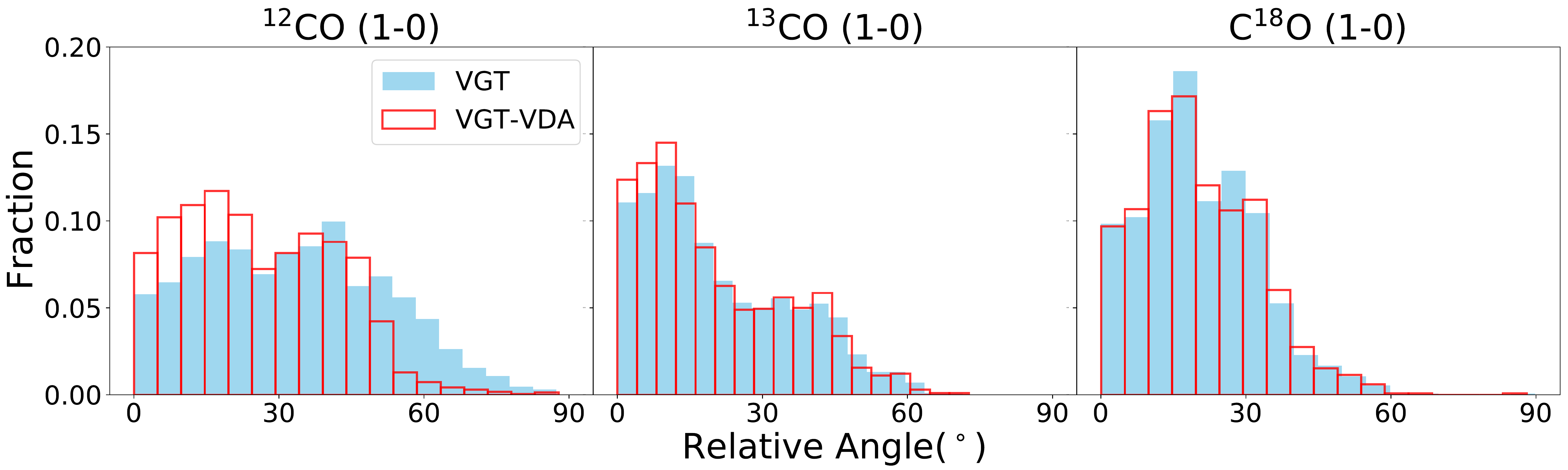}
    \caption{A histogram of the relative angle between the B-field measured from the Planck polarization and from the VGT method. 
    The blue columns show the relative angles for the VGT algorithm using the CO spectral line data. 
    The red columns show the relative angle distribution from the combined VGT-VDA algorithm. }
    \label{RA}
\end{figure*}

A detailed comparison between the magnetic fields from the Planck dust emission data with those from the VGT method may be achieved when setting a sub-block size for the VGT method results of 150$\times$150\,pixels, which makes VGT pseudo beam of 5$'$ the same as Planck 353\,GHz dust polarization (see Fig.\,\ref{VGT/P}). 
The VGT pixel size of the pseudo stokes maps will be re-gridded to 1.71$'$ and be equal to the pixel size of the Planck polarization. 
When considering the signal-to-noise ratio (SNR) greater than 3 times sigma for the B-field vectors for both the dust polarization and the spectral line data,  
the mean AM values are found to be about 0.34$\pm$0.01 for $^{12}$CO, 0.64$\pm$0.01 for $^{13}$CO and 0.70$\pm$0.01 for C$^{18}$O. 
This means that $^{13}$CO and C$^{18}$O trace magnetic field well when using VGT. 
$^{12}$CO, $^{13}$CO, and C$^{18}$O emission originate from different layers of cloud which has different critical density, $\sim$ 10$^2$, 10$^3$, and 10$^4$ cm$^{-3}$ \citep{1999ARA&A..37..311E,2015PASP..127..299S}, respectively.
Dust continuum at sub-mm wavelength traces dense region.
$^{13}$CO, and C$^{18}$O trace the dense gas whose origin of them is similar to that of dust emission at sub-mm wavelength.
Consequently, the high AM values observed for dense tracers, $^{13}$CO and C$^{18}$O, are to be expected since dense molecular tracers probe the dense molecular gas ($\sim$ 10$^{4}$ cm$^{-3}$).
$^{12}$CO traces more diffuse gas so that the velocity gradients are less aligned with the Planck polarization. 

In addition, the performance of the VDA methods for improving the VGT results may be tested.
Using the same method (see Eq.\,\ref{AMeq}) to compare the Planck and VGT-VDA B-field directions, the mean AM values from the three VGT-VDA rsults for $^{12}$CO,$^{13}$CO and C$^{18}$O  are 0.54$\pm$0.01, 0.66$\pm$0.01, and 0.71$\pm$0.01. 
The mean AM value from the VGT-VDA method for $^{12}$CO (AM =0.34) has been greatly improved compared with the VGT-only method (AM = 0.54). 
All mean AM values from the three VGT-VDA results are above 0.5.
It means that there is a smaller difference between the dust polarization results and the B-fields measured with VGT-VDA using the CO emissions. 
However, this improvement is insignificant for the $^{13}$CO and C$^{18}$O data (AM values improved by 0.01 $\sim$ 0.02), which may be explained by VDA only being effective in region with prevalent MHD turbulence (see \citealt{2021ApJ...910..161Y}). 
In the presence of dense gas, as for $^{13}$CO and C$^{18}$O, the VDA method does not seem to work well to improve the accuracy of the VGT method. 
Another possibility is that VDA relies on the high signal-to-noise ratio of the spectra \citep{2021ApJ...910..161Y}, while $^{13}$CO and C$^{18}$O data have a lower signal-to-noise ratio than $^{12}$CO.
It should be noted that the B-field measured by VGT for $^{12}$CO  in OMC-1 has distinct differences from the B-field measured by dust polarization, which we will discussed in detail in Zhao et al.(in prep).

Compared with VGT results from the different tracers, the AM (VGT-VDA) from $^{13}$CO and C$^{18}$O is above 0.65. 
The magnetic fields derived from the VGT-VDA method for $^{13}$CO and C$^{18}$O  are more similar to the B-fields derived from the dust polarization emission than that from the $^{12}$CO emission. The critical density for the optically thick tracer $^{12}$CO (1-0) emission is around 10$^{2}$\,cm$^{-3}$ as it traces the diffuse regions of the cloud. 
The Planck 353\,GHz dust polarized emission generally combined contributions from both the diffuse and the dense regions. 
The optical depth for $^{13}$CO and C$^{18}$O is generally lower than for $^{12}$CO. 
The pseudo magnetic field measured with VGT therefore samples regions with different critial densities corresponding molecular tracer. 
The Planck 353\,GHz dust polarization trace the magnetic field from optically thin and dense region. 
Since $^{13}$CO and C$^{18}$O sample similar regions, their VGT results would thus also be close to the local magnetic field inferred from the Planck 353\,GHz dust polarization.  

\subsection{The VDA effect on VGT}

To see whether the VDA method does improve the results of the VGT method in tracing the magnetic fields, the Planck polarization has been set as reference and the relative angle has been set as the absolute value of offset angle $\theta_r$ (see Eq\,.\ref{off}). 
Figure\,\ref{RA} shows three histograms of the relative alignment between the Planck polarization and the VGT results from $^{12}$CO, $^{13}$CO, and C$^{18}$O. 


The left panel of Fig.\,\ref{RA} shows that relative angles from the VGT-VDA method for $^{12}$CO are closer to zero degrees than those for the VGT method. 
The columns for the VGT-VDA method between 60$^\circ$ and 90$^\circ$ show a significant decrease compared with those of the VGT method, which indicates that the VDA algorithm significantly improves the accuracy of VGT tracing by down-shifting the B-field angles for $^{12}$CO. 
The middle panel of Fig.\,\ref{RA} for $^{13}$CO shows that between 60$^\circ$ and 90$^\circ$ the relative angle distribution is suppressed for both VGT and VGT-VDA methods. 
The VGT-VDA method results in a significant improvement from 0$^\circ$ to 30$^\circ$ over those for the VGT method. 
The mean AM value for the $^{13}$CO VGT-VDA method (0.67) has not improved compared with the VGT method (0.66; see Sect.\ref{result}). 
The right panel of Fig.\,\ref{RA} right panel shows that most of the relative angle for both VGT and VGT-VDA are in the 0$^\circ$ $\sim$ 30$^\circ$ range and that there is a small difference between the two methods.  
The middle and right panels suggest that the VGT-VDA method gives a small improvement over the VGT-only method for tracing the magnetic field when using the density tracers $^{13}$CO and C$^{18}$O.

\begin{figure*}
    \centering
    \includegraphics[width=18cm]{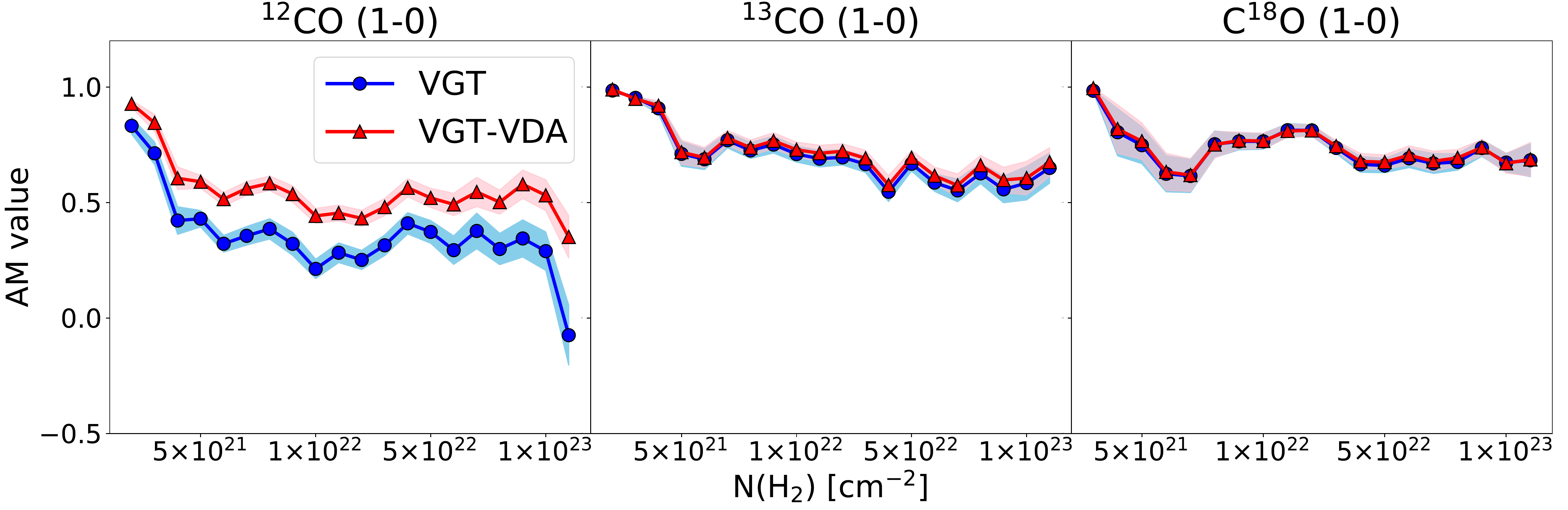}
    \caption{The Alignment Measure distribution versus N(H$_2$). 
    The blue line with circles is the AM value from only the VGT method. 
    The red line with triangles is the AM value from the combined VGT-VDA method.}
    \label{nhv}
\end{figure*}

\begin{table*}
\small
    \centering
    \caption{Physical parameters of the molecular clouds at Orion A}
    \begin{tabular}{c c c c c c c c c }
    \hline
	\hline
    Property & Units  & OMC-1 & OMC-2 & OMC-3 & OMC-4 & OMC-5 & L1641-N & NGC1999 \\
    \hline
    Area & pc$^2$ & 0.32 & 0.30 & 0.73 & 0.54 & 0.62 & 1.68 & 0.14  \\ 
    1D velocity dispersion  & km s$^{-1}$ & 3.29 & 2.06 & 2.01 & 2.99 & 2.69 & 3.04 & 2.85\\
    B-field angle dispersion & degree & 9.5 & 12.1 & 16.6 & 4.9 & 5.3 & 4.52 & 4.1  \\
    H$_2$ column density & $\times 10^{22}$\rm cm$^{-2}$ &  6.97  & 1.57 & 4.90  & 2.25  & 1.53  & 1.27 & 2.26 \\
    Effective radius& pc & 0.12 & 0.14 & 0.14 & 0.15 & 0.15 &  0.73 & 0.21  \\
    Effective length & pc  & 1.12 & 0.88 & 1.5 & 1.40 & 1.72 & 1.46 & 0.43 \\
    H$_2$ volume number density & $\times 10^{4}$ $\rm cm^{-3}$ & 12.02 & 2.07 & 6.47 & 2.16 & 2.08 & 0.28  & 1.71\\
    Volume mass density & $\times10^{-19}$ g cm$^{-3}$ & 5.64 & 1.80 & 0.97 & 1.48 & 0.98 &  0.13 & 0.80 \\
    Kinetic temperature & K & 31.6 & 19.6 & 25.4 & 13.3 & 9.1 & $\sim$ & $\sim$  \\
	Dust temperature & K & 30.2 & 20.5 & 21.1 & 19.2 & 15.9 & 15.7 & 15.7 \\
	Sound speed & km s$^{-1}$ & 0.33 & 0.26 & 0.30 & 0.21 & 0.18 & 0.23 & 0.23 \\
    Sonic Mach number & & 9.92 & 7.89 & 6.75 & 13.91 & 15.10 & 13.0 & 12.22 \\
    Alfv\'{e}n number && 1.51 & 1.80 & 1.50 & 1.33 & 1.41 & 1.61 & 1.79\\
    Magnetic field strength (DCF) & $\mu$G & 1114 & 310 & 161 & 1013 & 679 & 331 & 837 \\
    \it{B}$_{\rm pos}$ \rm (MM2) & $\mu$G & 1061 & 172 & 148 & 532 & 211 & 148 & 161  \\
    \hline
    \hline
    \end{tabular}
    \tablecomments{The location of these molecular clouds has been shown in Fig.\,\ref{fig2}). Details of parameters used for the calculated function have been shown in Table\,\ref{mAp}. 
    The parameter \emph{T} from c$_s$ is the kinetic temperature \emph{T}$_{\rm k}$ for OMC-1,\,2,\,3,\,4,\,5 \citep{2017ApJ...843...63F} and the dust temperature \emph{T}$_{\rm d}$ for other clouds.
    There are two models: the cylindrical model applied for OMC-1,\,2,\,3,\,4,\,5 and the spherical model applied for L\,1641-N and NGC\,1999. 
    Specific calculation formulas have been shown in Appendix\,\ref{A-A}. 
    The magnetic field strength (DCF) is calculated from the VGT dispersion and the DCF method (see \S\,\ref{V-DCF}). 
    The parameter B$_{\rm pos}$ (MM2) is the POS magnetic field strength measured by MM2 (see \S\,.\ref{DMA}). }
    \label{mgs}
\end{table*}



The alignment of the pseudo magnetic field from the VGT method depends on the local column density. 
The column density distribution in Orion A has been derived using SED fitting (\citealt{2013ApJ...763...55R,2013ApJ...777L..33P};see Fig.\,\ref{fig1}). 
Fig.\,\ref{nhv} shows that there is a correlation between column density and the AM value from both the VGT and VGT-VDA methods. 
The left panel of this Fig.\,\ref{nhv} shows that the AM values from the $^{12}$CO VGT-VDA procedure is always higher than the VGT values at different column densities. 
Removing the density contribution with VDA gives a more consistent relationship with the dust emission. 
The other two panels in Fig.\,\ref{nhv} show that using the VDA method gives a small improvement of AM values for $^{13}$CO and essentially no change for C$^{18}$O. 
The VDA effect is less at dense regions. 
All AM values from $^{13}$CO and C$^{18}$O are above 0.5 for all column densities, which makes 
them good tracers of the magnetic field by using the VGT method.

In regions with different column densities, the AM values from the three CO emissions show a consistent trend: the AM values decline with increasing column density. 
Orion\,A is an active star-forming region and has many complex and interesting structures such as filaments, bipolar outflows, shells, bubbles, and photo-eroded pillars \citep{2018ApJS..236...25K,2018A&A...609A..16T,2020ApJ...901...62L}. 
These regions are not simply dominated by either turbulence or self-gravity. 
More diffuse regions has weak star-forming activity in comparison with the denser regions and the motion traced by turbulence and velocity caustics would be more aligned with the local magnetic field.
This makes using VGT more conducive for tracing magnetic fields.

For all three CO lines, the AM values in Fig.\,\ref{nhv} show a downward trend for a column density in the range 3$\times$10$^{21}$ to 5$\times$10$^{21}$\,cm$^{-2}$. 
When the column density is above 5$\times$10$^{21}$\,cm$^{-2}$, the AM values do not drop drastically and remain near a stable value. 
The stable AM values from VGT-VDA for $^{12}$CO, $^{13}$CO and C$^{18}$O are around 0.5, 0.7 and 0.7. 
In those dense regions the 353\,GHz dust emission would be mostly optically thin and trace the inner projected magnetic field. 
$^{13}$CO and C$^{18}$O are also optically thin molecular tracers of the same inner structure in the molecular cloud, which would make their VGT results similar to the 353\,GHz dust polarization (at 850$\mu$m). 
In relative terms, $^{12}$CO is an optically thick tracer and traces the surface structure of clouds and is greatly affected by the density contribution in its velocity channel. 
The VDA removal of the density influence greatly improves the VGT accuracy and results in magnetic fields similar to those of the inner clouds. 


\subsection{Magnetic Field Strength derived from VGT}

\subsubsection{VGT dispersion with DCF method}\label{V-DCF}

The dust polarization results are more consistent with the VGT results from $^{13}$CO and C$^{18}$O than with those from $^{12}$CO. 
Even in high-density areas, the AM values for VGT from $^{13}$CO and C$^{18}$O are above 0.5. 
Since the area covered by the $^{13}$CO emission is much larger than that of C$^{18}$O, $^{13}$CO would be a better tracer of the magnetic field in Orion\,A using the VGT method. 
The VGT-VDA method from $^{13}$CO may then be used to calculate the magnetic field correlation parameter in Orion\,A. 
The resolution of the magnetic field from $^{13}$CO is close to 40$''$ (FWHM $\approx$ 0.07 pc).

Earlier the Davis-Chandrasekhar–Fermi (DCF; \citealt{1951PhRv...81..890D,1953ApJ...118..113C}) method has been used in Orion\,A to estimate the magnetic field strength defined as \citep{2004ApJ...600..279C,2017ApJ...846..122P}:
\begin{equation}
    \begin{aligned}
    \it{B}_{\rm pos} \approx \rm 9.3\sqrt{n(H_2)}\frac{\Delta \it{v}_{1D}}{\langle\sigma_{\rm \theta}\rangle} \,\mu G \,,
    \end{aligned}
\end{equation}
where $\Delta v_{1D}$  is the line width (FWHM) of the molecular line in km s$^{-1}$, $\sigma_{\rm\theta}$ is the angular dispersion from dust polarization or VGT methods in degrees, and n(H$_2$) is the Hydrogen volume density in cm$^{-3}$. 
The parameters $\langle\sigma_{\rm \theta}\rangle$ and $\it{B}_{\rm pos}$ are the mean angular dispersion in Orion\,A and the mean plane-of-the-sky magnetic field strength at this region. 
However, the DCF method originally did not consider self-gravity and sub-regions of Orion\,A are the dense region, include OMC-1, OMC-2, OMC-3, OMC-4, OMC-5, L\,1641-N, and NGC\,1999, where is gravitationally collapse \citep{2017A&A...602L...2H}. 
The magnetic field strength estimated with the VGT and DCF methods should be compared and the errors evaluated.

OMC-1, OMC-2, OMC-3 OMC-4, and OMC-5 are located along the large Integral Shaped Filament (ISF). 
The shape of these filaments is roughly like a long cylinder. 
Two other components, L\,1641-N and NGC\,1999, are dense clumps in Orion\,A and have a morphological structure that is simply spherical. 
Using these two simple geometric models (cylinder and spherical, see Appendix.\ref{A-A}), the physical parameters in these clouds have been calculated and are presented in Table\,\ref{mgs}. 
The functions to calculate these parameters has been shown in Table\,\ref{mAp}. 
Because of the 1D velocity dispersion $\sigma_{v,1D}$ that includes a turbulence velocity and a shear velocity, the magnetic field strength $\it{B}_{\rm pos}$ could be overestimated.

\begin{figure}
    \centering
    \includegraphics[width = 8.6cm]{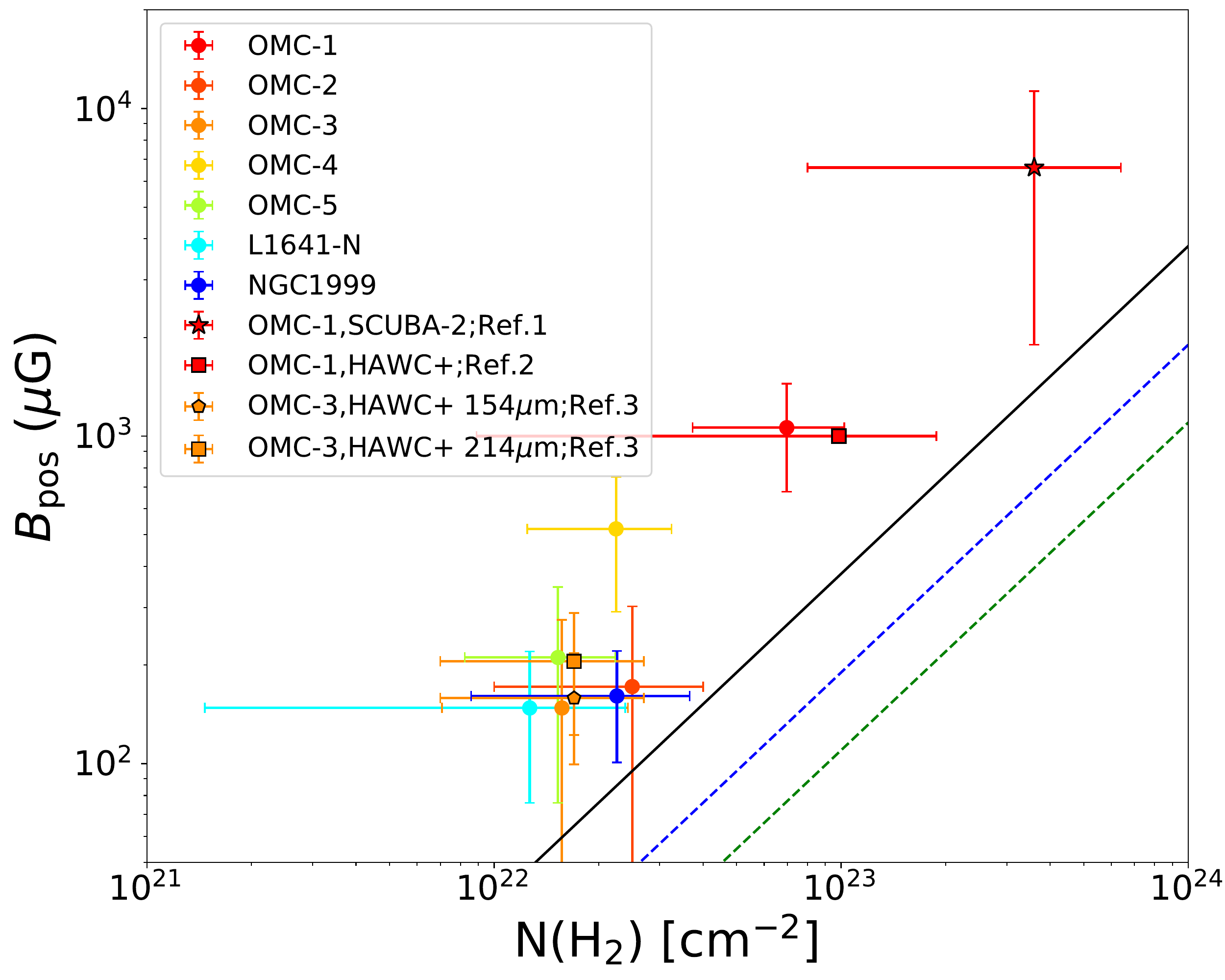}
    \caption{The relation between the magnetic field strength in the plane of the sky($\it{B}_{\rm pos}$) and the \rm H$_2$ column density (N(H$_2$)). 
    The black line is a critical condition for the mass to flux ratio($\lambda$ = 1). 
    The green dashed line is an empirical relationship between the magnetic field strength and the column density\citep{2021arXiv211105836L}. 
    The magnetic field strength of the molecular clouds is distributed along this line. 
    The red star is the OMC-1 magnetic field strength observed by JCMT\citep{2017ApJ...846..122P}.
    The Red square is $\it{B}_{\rm pos}$ observed by HAWC+ \citep{2019ApJ...872..187C}. 
    The Orange square and pentagon are $\it{B}_{\rm pos}$ by HAWC+ for bands D and E \citep{2021arXiv211110252Z}. 
    The colorful circles show the magnetic field strength measured by the MM2 methods (see Sect.\,\ref{DMA}).}
    \label{Bnh}
\end{figure}

\subsubsection{Two Mach Numbers Analysis}\label{DMA}

A basic assumption of the classical DCF techniques is that the observed fluctuations in the molecular medium result from Alfvén waves (see \S~\ref{V-DCF}; \citealt{1951PhRv...81..890D,1953ApJ...118..113C}). 
Self-gravity in the region causes additional fluctuations in the magnetic field and the turbulence, which would make the classical DCF method incomplete. 
A new technique called Two Mach Numbers method (MM2), which $\it{B}_{\rm pos}$ calculated by Alfvén and sonic Mach number, provides an alternative for measuring the magnetic field strength when external shear and self-gravity distorts the magnetic fields. 
The associations of the MM2 method were analytically justified in \cite{2020arXiv200207996L}.
The velocity gradients and also the magnetic fields are a function of the Alfvén Mach number $M_A$ within a sub-block \citep{2020arXiv200207996L,2021ApJ...912....2H}. 
The dispersion relation in the direction of the velocity gradient will show a power-law alignment with the $M_A$, which is a so-called "top-to-bottom" ratio of the distribution of the fine channel velocity gradients (VChGs). 
The Alfven Mach number $M_A$ may be calculated as:
\begin{equation}
	\label{eq.tb}
	\begin{aligned}
		{\rm M_A}&\approx1.6 (T_v/B_v)^{\frac{1}{-0.60\pm0.13}}, for\,{\rm M_A}\le1 \,, \\
		{\rm M_A}&\approx7.0 (T_v/B_v)^{\frac{1}{-0.21\pm0.02}}, for\,{\rm M_A}\textgreater 1 \,, \\
	\end{aligned}
\end{equation}
where $T_v$ denotes the maximum value of the fitted histogram of the velocity gradient orientation, while $B_v$ is the minimum value. 
And with the knowledge of two Mach number $M_{\rm S}$ amd $M_{\rm A}$, this new technique, MM2 \citep{2020arXiv200207996L}, may be used to calculate the POS magnetic field strength as:
\begin{equation}
    B=\Omega c_s\sqrt{4\pi\rho_0}\,{\rm M_S M_A^{-1}},
\end{equation}
where the $\Omega$ is a geometrical factor ($\Omega$ = 1), c$_s$ is the sound speed, $\rho_0$ is the volume mass density in units of g\,cm$^{-3}$, and $M_S$ and $M_A$ are the sonic Mach number and the Alfvén Mach number. 
The $\it{B}_{\rm pos}$ derived by $^{13}$CO MM2 is similar to the $\it{B}_{\rm pos}$ obtained from the VGT dispersion with DCF method in most sub-clouds (see Table\,\ref{mgs}), except for the case for NGC\,1999. 
The multi-velocity components of the $^{13}$CO spectral line emission could further increase the velocity dispersion and also $B_{\rm pos-DCF}$.

Figure\,\ref{Bnh} displays the magnetic field strength (measured by the MM2 mothed) distribution with H$_2$ column density. 
The different colors indicate the results for different molecular clouds. 
The circles are magnetic field strengths estimated by the $^{13}$CO VGT method. 
The magnetic field strength at OMC-1 has been estimated from the dust polarization \citep{2017ApJ...846..122P,2019ApJ...872..187C}. 
The B-field strength $B\rm _{pos}$ are 6.6\,mG (sub-mm;JCMT) and 0.9$\sim$1\,mG (Far-IR; HAWC+). 
The OMC-1 $B\rm _{pos}$ measured by the MM2 method for $^{13}$CO is around 1\,mG, which is similar to the dust polarized values. 
In OMC-3, $B_{pos}$ measured by MM2 is 148\,$\mu$G. 
The $B\rm _{pos}$ values derived by HAWC+(154\,$\mu$m and 214\,$\mu$m) are 158.6 and 205.4\,$\mu$G. 
The $B\rm _{pos}$ value from MM2 and dust polarization is similar to $B\rm _{pos}$ from the dust polarization. 
The $\it{B}_{\rm pos}$ value from the VGT dispersion and the DCF is larger than obtained from MM2. 
In the self-gravity region, the distorted magnetic field and the turbulence cause the estimated magnetic field strength with DCF to become larger. 
The MM2 method may provide another possible method for measuring the magnetic field strength in self-gravity regions.

\section{Summary}\label{sum}





The objective of this work is to measure the magnetic field in Orion\,A with the VGT and VGT-VDA methods by multiple CO spectra emission.
For this purpose, the VGT and VGT-VDA methods have been applied to determine the magnetic field structures in the filament structure of Orion\,A using the $^{12}$CO, $^{13}$CO, and C$^{18}$O\,(1-0) emission profiles with a spatial resolution of $\sim$ 0.07 pc. 
It has been found that the VGT-VDA method has a great accuracy in tracing the magnetic field at small scales and shows strong agreement with the larger scale field structures determined from the Planck dust polarization data.
In addition, the MM2 method would be an alternative to to estimate the POS magnetic field strength in regions where self-gravity plays a role.
Further results are the following:


1. The magnetic field morphology measured with the VGT method demonstrates east-west structural features in Orion\,A. The magnetic field orientation is mainly perpendicular to the direction of the Integral Shaped Filament. 

2. On the whole, the B-field morphology measured with VGT for $^{13}$CO is similar to that of C$^{18}$O. In dense regions, the orientations of the magnetic field derived by VGT for $^{12}$CO are comparable to those of $^{13}$CO and C$^{18}$O. In some relatively diffuse areas, the magnetic field orientations derived for $^{12}$CO are different from those of $^{13}$CO and C$^{18}$O.


3. In dense regions, the B-fields measured using VGT method for $^{13}$CO, C$^{18}$O and the VGT-VDA method for  $^{12}$CO, $^{13}$CO, C$^{18}$O are in agreement with those derived from the Planck 353\,GHz dust polarization at the same scale ($\sim$ 0.55\,pc). 
The AM values for these are 0.66, 0.70, 0.54, 0.66, and 0.71, respectively, and they are all over 0.5. 
This would indicate that the magnetic field measured with VGT is similar to that of dust polarization.



4. The VDA method can improve the accuracy of the VGT to trace magnetic fields by separating velocity and density contribution, specially for $^{12}$CO (eg, AM values from VGT and VGT-VDA method for $^{12}$CO are 0.33 and 0.54). 
In dense regions with N(H$_2$) $\textgreater$ 3 $\times$ 10$^{22}$ cm$^{-2}$, the VDA method does not significantly improve the accuracy of VGT for $^{13}$CO and C$^{18}$O. 
Additional corrections may be needed for the VDA method for tracers of dense regions.
The improved method, VGT-VDA, can provide a higher accuracy to trace magnetic field.



5. A new technique, MM2, has been applied for the $^{13}$CO data in Orion\,A to measure the magnetic field strength. The POS B$_{\rm pos}$ values for OMC-1, OMC-2, OMC-3, OMC-4, OMC-5, L\,1641-N and NGC\,1999 are 1061, 172, 148, 532, 211, 148, and 161\,$\mu G$, repectively, which is consistent with previous results obtained from dust polarization at far-infrared and submillimeter wavelengths. 

Plans are to continue the comparison of methods for measuring the magnetic field in more sources to test the VGT method in complex star formation regions.


\software{Julia \citep{2012arXiv1209.5145B}, \texttt{python, Ipython} \citep{2007CSE.....9c..21P}, \texttt{numpy} \citep{2011CSE....13b..22V}, \texttt{matplotlib} \citep{2007CSE.....9...90H},  \texttt{astropy} \citep{2013A&A...558A..33A}, \texttt{RadFil} \citep{2018ApJ...864..152Z}}

\section*{Acknowledgements}
The authors thank the anonymous referee for helpful comments.
The authors thank Dr. Tao-Chung Ching for providing codes of dust polarization and helpful comments. 
This work was mainly funded by the National Natural Science foundation of China (NSFC) under grant No.\,11973076. 
Partial funding was obtained from the NSFC under grant Nos.\,11903070, 12173075, and 12103082, 
the Natural Science Foundation of Xinjiang Uygur Autonomous Region under grant No.\,2022D01E06, the Heaven Lake Hundred Talent Program of Xinjiang Uygur Autonomous Region of China, 
the Tianshan Innovation Team Plan of Xinjiang Uygur Autonomous Region (2022D14020), 
the Youth Innovation Promotion Association CAS, 
the CAS Light of West China Program under grant No.\,2020-XBQNXZ-017, 
and the Project of Xinjiang Uygur Autonomous Region of China for Flexibly Fetching in Upscale Talents. 
A.L. and Y.H. acknowledge the support of the NASA ATP 80NSSC20K0542 and NASA TCAN 144AAG1967. 
W.A.B has been funded by Chinese Academy of Sciences President’s International Fellowship Initiative grant No.\,2021VMA0008. 
This research has made use of data from the Herschel Gould Belt survey (HGBS) project (\url{http://gouldbelt-herschel.cea.fr}). 
The HGBS is a Herschel Key Programme jointly carried out by SPIRE Specialist Astronomy Group 3 (SAG 3), scientists of several institutes in the PACS Consortium (CEA Saclay, INAF-IFSI Rome and INAF-Arcetri, KU Leuven, MPIA Heidelberg), and scientists of the Herschel Science Center (HSC).

\bibliography{reference}

\appendix

\section{Estimated Physical parameters}\label{A-A}

There are two models to estimate the physical parameter of the clouds in Orion A. A filament is assumed to be a long uniform cylinder and the clump is a uniform sphere \citep{2000MNRAS.311...85F}. These computational formulas have been shown in table.\ref{mAp}.

\begin{table*}[h]
    \centering
    \caption{Physical parameters of different model}
    \begin{tabular}{  c  c  c  c  }
    \hline
	\hline
    Propertty & Symbol & cylindrical Model & spherical Model\\
    \hline
    Area & $A$ &  \\ 
    1D velocity dispersion & $\sigma_{v,1D}$  \\
    B-field angle dispersion & $\sigma_{\theta}$   \\
    H$_2$ column density (\rm cm$^{-2}$) & $N_0$  \\
    Effective radius & $R$  & 0.5 times of filament width & $\sqrt{A/\pi}$ \\
    Effective Length(or diameter) &  $L$   & filament length & 2$R$\\
    H$_2$ volume number density & $n_0$ & 2N$_0/\pi$R & $N_0$/$L$ \\
	Sound speed & $c_{\rm s}$ & $\sqrt{k_{\rm B}T/\mu_pm_{\rm H}}$ & $\sim$\\
	Sonic Mach number & $M_{\rm S}$ & $\sigma_{v,1D}$/c$_{\rm s}$ & $\sim$ \\
	Alfv\'{e}n Mach number & $M_{\rm A}$ & MM2 (see \S\,\ref{DMA}) & $\sim$ \\
    \hline
    \hline
    \end{tabular}
    \tablecomments{Mark $\sim$ means that it is the same as the previous formula. Length and width of filaments were identified by Python Package: RadFil \citep{2018ApJ...864..152Z}. }
    \label{mAp}
\end{table*}

\end{document}